\documentclass[12pt, a4paper]{article}

\usepackage[utf8]{inputenc}
\usepackage{amsmath}
\usepackage{amsfonts}
\usepackage{amssymb}
\usepackage{amsthm}
\usepackage{amsbsy}
\usepackage{enumerate}
\usepackage{fancyhdr}
\usepackage{tikz}
\usepackage{hyperref}
\usepackage{url}
\usepackage{bm}
\usepackage{enumitem}
\usepackage{algorithm}
\usepackage{algpseudocode}

\usepackage{thm-restate}

\usetikzlibrary{babel}
\newtheorem*{thm*}{Theorem}
\newtheorem{theorem}{Theorem}

\newtheorem{lemma}[theorem]{Lemma}
\newtheorem{corollary}[theorem]{Corollary}

\theoremstyle{definition}

\newcommand{\FQW}{\mathcal{FQW}}
\newcommand{\F}{\mathcal{F}}
\newcommand{\Q}{\mathcal{Q}}
\newcommand{\W}{\mathcal{W}}
\newcommand{\R}{\mathcal{R}}
\newcommand{\rtop}{r_{T}}
\newcommand{\rleft}{r_{L}}
\newcommand{\rbot}{r_{B}}
\newcommand{\LM}{{LM}}
\newcommand{\StripPacking}{\textsc{Strip Packing}}
\newcommand{\SquareStripPacking}{\textsc{Square Strip Packing}}
\newcommand{\BinPacking}{\textsc{Bin Packing}}
\newcommand{\LoadBalancing}{\textsc{Load Balancing}}
\newcommand{\BottomLeftAlgorithm}{\textsc{Bottom-Left Algorithm}}
\newcommand{\BL}{\textsc{BL}}

\newcommand{\MultipleStripPacking}{\textsc{Multiple Strip Packing}}

\parskip=\smallskipamount
\let\epsilon=\varepsilon

\usetikzlibrary{decorations.pathreplacing}

\begin{document}

\title{A 13/6-Approximation for Strip Packing via the Bottom-Left Algorithm}
\author{Stefan Hougardy\thanks{Research Institute for Discrete Mathematics and Hausdorff Center for Mathematics, University of Bonn, Germany (hougardy@dm.uni-bonn.de) funded by the Deutsche Forschungsgemeinschaft (DFG, German Research Foundation) under Germany's Excellence Strategy -- EXC-2047/1 -- 390685813}
\and Bart Zondervan\thanks{Faculty of Mathematics and Computer Science, University of Bremen, Germany, (bart.zondervan@uni-bremen.de)}}
\date{}
\maketitle

\definecolor{green}{rgb}{0, 0.7, 0}%
\definecolor{blue}{rgb}{0, 0, 0.95}%
\definecolor{mybrown}{rgb}{0.7, 0.4, 0}

%Fillstyle for all rectangles
\tikzset{myfill/.style ={fill=blue!20!white, draw=black}}
\tikzset{mygreenfill/.style  ={green!40!white, draw=black}}
\tikzset{mybrownfill/.style  ={mybrown!40!white, draw=black}}

% START -------------------------------------------------------------------------------------------

\begin{abstract}
In the \StripPacking\ problem, we are given a vertical strip of fixed width 
and unbounded height, along with a set of axis‑parallel rectangles. 
The task is to place all rectangles within the strip, without overlaps, 
while minimizing the height of the packing. This problem is known to be NP-hard.
The \BottomLeftAlgorithm\ is a simple and widely used heuristic for \StripPacking. 
Given a fixed order of the rectangles, it places them one by one, always choosing the lowest feasible position in the strip and, in case of ties, the leftmost one. 
Baker, Coffman, and Rivest proved 
in 1980 that the \BottomLeftAlgorithm\ has approximation ratio~$3$ if the rectangles are sorted by decreasing width~\cite{BCR1980}. 
For the past 45 years, no alternative ordering has been found that improves this bound. We introduce a new rectangle ordering and show that with this ordering the \BottomLeftAlgorithm\ achieves a $13/6$
approximation for the \StripPacking\ problem.
\end{abstract}

\newpage
\section{Introduction}
The \StripPacking\ problem is a fundamental problem in combinatorial optimization, with applications in cutting-stock manufacturing, VLSI design, and scheduling.
In this problem, rectangles must be packed orthogonally into a strip of fixed width, minimizing the total height used.

The \StripPacking\ problem is a generalization of both \BinPacking\ and \LoadBalancing\ ($P||C_\text{max}$), which implies that the problem is NP-hard~\cite{Karp} and that there cannot exist a polynomial-time algorithm with absolute approximation ratio~$3/2-\epsilon$, unless~P=NP.
Additionally, the problem is known to be strongly NP-hard~\cite{Garey78}.

These hardness results provide strong motivation for the development of efficient approximation algorithms.
The earliest of such algorithms for \StripPacking\ is the \BottomLeftAlgorithm\ (called the \BL\ algorithm for short), introduced in~$1980$ by Baker, Coffman, and Rivest~\cite{BCR1980}. 
This algorithm is conceptually simple yet powerful: it processes the input rectangles in a predetermined order, placing each at the lowest possible position in the strip and breaking ties by choosing the leftmost position. 
Despite its simplicity, the \BL\ algorithm laid the foundation for decades of research into geometric packing heuristics and many open questions surrounding the algorithm remain.
In their pioneering work, Baker et al.\cite{BCR1980} proved that the \BL\ algorithm achieves an absolute approximation ratio of~$3$, and no better bound for this algorithm has been established since.
Furthermore, Chazelle~\cite{Chazelle1983, Mich2025} established a quadratic runtime implementation of the \BL\ algorithm.

Subsequent research on absolute approximation algorithms for~\StripPacking\ 
has led to significant advances~\cite{Coffman1980, Sleator19802, Schiermeyer1994, Steinberg1997, Harren2009improved19396}.
In~$2014$, Harren, Jansen, Pr\"adel, and van Stee~\cite{harren2014} presented an algorithm achieving the current best-known absolute approximation ratio of~$5/3+\epsilon$.
For a special case in which all rectangles are skewed (that is, each rectangle has either width less than a~$\delta$-fraction of the strip width or height less than a~$\delta$-fraction of the optimal packing height), G\'{a}lvez, Grandoni, Jabal Ameli, Jansen, Khan, and Rau~\cite{galvez2023} further designed an almost tight~$(3/2+\epsilon)$-approximation.
While these results represent major theoretical progress, the proposed algorithms are intricate and appear to have limited applicability in practical settings.

In contrast, the \BottomLeftAlgorithm\ stands out for its conceptual simplicity and its widespread use in real-world scenarios~\cite{jakobs1996genetic, rolich2016bottom, fang2024new}.

As the performance of the~\BL\ algorithm strongly depends on the predetermined order of the rectangles, it is tempting to think that among the~$n!$ possible orderings of~$n$ rectangles, at least  one of them produces an optimal packing.
Unfortunately, this is not the case: even under the best possible ordering, the absolute approximation ratio of the \BL\ algorithm cannot surpass~$4/3-\epsilon$~\cite{hougardy2024bottom}, a fact already foreshadowed in earlier works~\cite{BCR1980,brown:1980}.
Nevertheless, the~\BL\ algorithm is a~$3$-approximation when rectangles are ordered by decreasing width (using an arbitrary order for rectangles with the same width)~\cite{BCR1980}.

Although several orderings of rectangles have been explored over the past~$45$ years~\cite{BCR1980, fekete2014online, Zond2023}, no ordering has been found that improves this~$3$-approxima\-tion.
To illustrate this, we present in Table~\ref{table:BL-apx-ratios} the known approximation ratios achieved by the \BL\ algorithm under different ordering strategies, for both the special case of \SquareStripPacking\ (all rectangles are squares) and for general \StripPacking.
\begin{table}
    \caption{Upper bounds on the approximation ratio of the \BottomLeftAlgorithm\ for different orderings of squares and rectangles. Results with a star have a matching lower bound.}
\centering
    \begin{tabular}{p{3.5cm}p{4.5cm}p{4.4cm}}
     \hline
     \textbf{Ordering} & \textbf{Square \StripPacking} & \textbf{\StripPacking} \\
     \hline
     Decreasing width & $2^\star$~\cite{BCR1980}  & $3^\star$~\cite{BCR1980} \\
     Decreasing height & $2^\star$ (as above) & unbounded~\cite{BCR1980} \\
     Decreasing area & $2^\star$ (as above)  & unbounded (Appendix) \\
     Increasing width & $3$~\cite{Zond2023}  & unbounded~\cite{BCR1980} \\
     Increasing height & $3$ (as above)  & $\geq 3$~\cite{Zond2023} \\
     Increasing area & $3$ (as above)  & unbounded (Appendix) \\
     Last-Row-Full & $2^\star$~\cite{Zond2023}   & undefined \\
     Arbitrary & $16$~\cite{fekete2014online, Zond2023} & unbounded~\cite{BCR1980} \\
     $\FQW$ & $2^\star$ (Section~\ref{sec:square_FQW}) & $\frac{13}{6} \approx \textbf{2.167}$ (Theorem~\ref{thm:main_13over6_theorem})\\
     \hline
    \end{tabular}
    \label{table:BL-apx-ratios}
\end{table}
The main contribution of this paper is to show that, for a carefully chosen ordering, which we call $\FQW$, the~\BL\ algorithm attains an approximation ratio of~$13/6$, improving the best known bound for this classic algorithm.

\begin{restatable}{theorem}{thmmain}
    The \BottomLeftAlgorithm\ for \StripPacking\ has absolute approximation ratio~$13/6$.
    \label{thm:main_13over6_theorem}
\end{restatable}

This result substantially narrows the gap between the known lower bound of~$4/3-\epsilon$ and 
the old upper bound of~3.
Importantly, it leaves open the intriguing possibility that the \BL\ algorithm could eventually outperform the current state-of-the-art~$(5/3+\epsilon)$-approximation, which serves as a primary motivation for our work.

\subparagraph*{Related Work.}
Kenyon and R\'emila~\cite{KR2000} proved that there exists an \emph{asymptotic} fully polynomial-time approximation scheme for \StripPacking. 
Their result has been sharpened in~\cite{JS2009,Svi2012}.
Furthermore, Harren and van Stee \cite{Harren2009improved19396} provided a polynomial-time approximation scheme for \StripPacking\ where rotations of rectangles by 90 degrees are permitted. This also implies a PTAS for~\SquareStripPacking\ (all rectangles are squares).

Beyond classical \StripPacking, the \BottomLeftAlgorithm\ has also been studied in other contexts.
In \MultipleStripPacking, where there are several strips of different widths, Zhuk~\cite{zhuk2006approximate} constructs a~$10$-approximation that combines an online algorithm for assigning the rectangles to strips, with the~\BL\ algorithm using decreasing-width order to place the rectangles within the strip.
Furthermore, Fekete, Kamphans, and Schweer~\cite{fekete2014online} considered the online variant of \SquareStripPacking\ with Tetris and gravity constraints, proving a competitive ratio of~$3.5$. 
The methods developed in this paper may also be useful in these contexts, offering the potential for improved bounds or deeper understanding of the \BL\ algorithm in its various extensions.

Additionally, the \BottomLeftAlgorithm\ fits within the framework of fixed priority algorithms as pioneered by Borodin, Nielsen and Rackoff~\cite{BorodinNR03}, since it irrevocably places rectangles in a predetermined order without knowledge of future rectangles.
Hence granting the algorithm the power to choose the ordering of the rectangles from an online instance yields constant competitiveness.

\section{Preliminaries}
A \StripPacking\ instance~$\mathcal{I} = (\R,W)$ consists of a vertical strip of fixed width~$W$ and infinite height, together with a set~$\R$ of~$n$ closed axis-aligned rectangles.
Each rectangle~$r$ has a given width~$w_r$ and height~$h_r$.
The maximum height of a rectangle in~$\R$ is denoted by~$h_\text{max}(\R)$ or simply by~$h_\text{max}$ if the set~$\R$ is clear from the context.
We assume that each rectangle~$r$ of the instance fits into the strip, i.e.,~$w_r\leq W$.

A \textit{packing} of the rectangles~$\R$ into the strip is defined by specifying the lower left coordinate~$(x_r,y_r)$ of each rectangle~$r\in\R$ which we call the \textit{position} of the rectangle $r$.
The packing is \textit{feasible} if all rectangles lie within the strip and no two rectangles overlap within their interior, i.e., for each~$r\neq r'\in \R$ the packing satisfies the two conditions:
\begin{align*}
    x_r\geq 0, \quad \quad x_r+w_r \leq W, \quad \quad y_r &\geq 0, \\
    (x_r,x_r+w_r) \times (y_r,y_r+h_r) \cap (x_{r'},x_{r'}+w_{r'})\times (y_{r'},y_{r'}+h_{r'}) &= \varnothing.
\end{align*}
Define the \textit{height} of a feasible packing as~$\max\{ y_r + h_r : r \in \R\}$.
The goal of~\StripPacking\ is to compute a feasible packing of minimum height for the given instance~$\mathcal{I}$.
Denote this value by~$h_\text{OPT}(\mathcal{I})$ or simply by~$h_\text{OPT}$.
We do not allow rotation of rectangles in this paper.
%The \SquareStripPacking\ problem is the special case of~\StripPacking\ where all rectangles are squares.

\subsection{The \BottomLeftAlgorithm}\label{sec:BL_algo}
The \BottomLeftAlgorithm\ is a simple and widely used heuristic for \StripPacking.
The algorithm was introduced by Baker et al.~\cite{BCR1980}.
Given a~\StripPacking\ instance~$(\R,W)$ together with an ordering of the rectangles~$r_1,\dots,r_n$,
the \BL\ algorithm places each rectangle in the specified order at the lowest available position in the strip, choosing the leftmost such position in case of a tie.
More formally, the \BL\ algorithm places the rectangle~$r_1$ at position~$(0,0)$, which is a feasible packing of the first rectangle.
Next, assuming that the \BL\ algorithm has obtained a feasible packing of the first~$i-1$ rectangles into the strip, it chooses a position~$(x_i,y_i)$ that results in a feasible packing for the first~$i$ rectangles, such that~$(y_i,x_i)$ is lexicographically minimum among all possible choices for the position~$(x_i,y_i)$.
Observe that this~\BL\ packing heavily depends on the ordering of the rectangles, which is part of the input rather than being computed by the algorithm.
To be precise, for a~\StripPacking\ instance~$\mathcal{I}=(\mathcal{R},W)$ and a sorting algorithm~$\mathcal{A}$ of the rectangles in~$\mathcal{I}$, denote by~$\mathcal{I}_\mathcal{A}$ the ordered instance where the rectangles are sorted according to~$\mathcal{A}$.
We call the packing obtained by the \BL\ algorithm a \textit{\BL\ packing} and denote it by~$\text{BL}(\mathcal{I}_\mathcal{A})$ for an ordered instance~$\mathcal{I}_\mathcal{A}$.
The height of a \BL\ packing on an instance~$\mathcal{I}_\mathcal{A}$ is denoted by~$h_\text{BL}(\mathcal{I}_\mathcal{A})$.
If the ordering is clear from the context, then we might drop the additional notation.

An important property of the \BL\ algorithm is that for every rectangle~$r_i\in \R$ not touching the left strip boundary, there is a rectangle~$r_j$ whose right face touches the left face of~$r_i$ and that is before~$r_i$ in the ordering, i.e.,~$j<i$ (as otherwise $r_i$ could be placed further to the left).
We call the rectangle~$r_j$ a \textit{left supporter} of~$r_i$.
A left supporter prevents the rectangle~$r_i$ from being placed more to the left.
Analogously, every rectangle that does not touch the bottom strip boundary has a \textit{bottom supporter}.

The \textit{(absolute) approximation ratio} achieved by the \BL\ algorithm on an instance~$\mathcal{I}$ ordered by~$\mathcal{A}$ is defined as the ratio~$h_\text{BL}(\mathcal{I}_\mathcal{A})/h_\text{OPT}(\mathcal{I})$.

\subsection{The Horizontal Strip Partition} \label{sec:horizontal_strip_partition}
For our analysis of the \BL\ algorithm we construct a partition of the strip based on the \BL\ packing of the rectangles.
This partition might be of interest independent of our results, and has already been used by Baker et al.~\cite{BCR1980}, although they did not explicitly mention it.
The main idea is to partition the strip into horizontal regions, where a region is the space under the bottom face of a rectangle and above the bottom faces of all rectangles placed before by the \BL\ algorithm.
The crucial property of this partition is that a region does not contain the bottom faces of any rectangle placed prior to placing the first rectangle above the region, and hence each unoccupied gap in this region has width strictly less than the width of this first rectangle that is placed above it, as otherwise the \BL\ algorithm could have placed the rectangle at a lower position.
We make this formal in Lemma~\ref{lemma:gap_width}.

Formally, let~$\R = \{r_1,\dots,r_n\}$ such that the \BL\ algorithm places~$r_i$ before~$r_j$ if~$i < j$.
We define the \textit{horizontal strip partition} of a \BL\ packing as the partition of the strip~$[0,W]\times[0,h_\text{BL}(\R,W)]$ into the regions
\begin{align*}
    H_i ~=~ \begin{cases}
       [0,W]\times [\max\{y_{r_1},\dots,y_{r_{i-1}}\}, y_{r_{i}}) &\quad\text{if } 1\leq i \leq n, \\
       [0,W]\times [\max\{y_{r_1},\dots,y_{r_n}\}, h_\text{BL}] &\quad\text{if } i=n+1.
    \end{cases}
\end{align*}
Here~$[x,x) = \varnothing$ and~$\max\varnothing = 0$.
Thus,~$H_i$ is the space below the bottom face of~$r_{i}$ and above the bottom faces of~$r_1,\dots,r_{i-1}$ and~$H_{n+1}$ is all the space above the highest bottom face. See Figure~\ref{fig:FQW-BL_example} for an example.

\subsection{Covering Proper Horizontal Lines}\label{sec:cov_hor_lines}
A key component in our analysis of the approximation ratio of the \BL\ algorithm is to bound  the total area occupied by rectangles in the \BL\ packing in terms of~$W\cdot h_\text{OPT}$.
Baker et al.~\cite{BCR1980} show that when the \BL\ algorithm places the rectangles in order of decreasing width, then each region~$H_i$ is at least half occupied by rectangles for all~$1\leq i \leq n$.
Moreover, as the height of~$H_{n+1}$ is at most~$h_\text{OPT}$, this implies that the \BL\ algorithm is a~$3$-approximation.
Using similar arguments, we will derive a slightly stronger result in Lemmas~\ref{lemma:gap_width} and~\ref{lemma:1/3_occupied_line}.

For this, we want to analyze the fraction of a horizontal line in the strip that is occupied by rectangles.
We call a horizontal line \textit{proper} if it does not intersect the top or bottom face of any rectangle.
We do not need to consider lines that are not proper, because their total area is zero.
A proper horizontal line alternates between parts that are occupied by rectangles and parts that are unoccupied.
An \textit{unoccupied gap} is a maximal connected unoccupied part on the line.
\begin{lemma}\label{lemma:gap_width}
    Let~$i\in \{1,\dots,n\}$ and~$\ell$ be a proper horizontal line in the region~$H_i$.
    Then each unoccupied gap in~$\ell$ has width strictly less than~$w_{r_{i}}$.
\end{lemma}
\begin{proof}
    By definition of~$H_i$, all rectangles placed prior to~$r_i$ have their bottom face below~$H_i$.
    Hence, if an unoccupied gap in~$\ell$ has size at least~$w_{r_{i}}$ when~$r_i$ is placed, then
    this contradicts the \BL\ algorithm, because the gap is a lower position where we can place~$r_i$. 
\end{proof}

\begin{lemma}\label{lemma:1/3_occupied_line}
    Let~$i\in \{1,\dots,n\}$.
    Suppose there exists a proper horizontal line~$\ell$ in the region~$H_i$ that intersects~$k$ rectangles just before~$r_i$ is placed.
    If each of these~$k$ rectangles has width at least~$w_{r_i}$, then any proper horizontal line in~$H_i$ is at least~$k/(2k+1)$ occupied.
    In particular, it is at least~$1/2$ occupied if an endpoint of~$\ell$ is occupied by a rectangle prior to placing~$r_i$. 
\end{lemma}
\begin{proof}
    We consider the iteration of the \BL\ algorithm in which rectangle~$r_{i}$ is placed. 
    By Lemma~\ref{lemma:gap_width}, in this iteration, each occupied part of~$\ell$ is at least as wide as each unoccupied gap, because the rectangles intersecting~$\ell$ have width at least~$w_{r_i}$ by assumption.
    If the line contains~$k\ge 1$ occupied parts in this iteration then it can contain at most~$k+1$ unoccupied parts. 
    Thus, at least a $k/(2k+1)$ fraction of the line is occupied. 
    If the leftmost or rightmost part of the line is occupied by a rectangle, then there are at most as many unoccupied parts as there are occupied parts. 
    Hence in this case at least a~$k/(2k) = 1/2$ fraction of the line is occupied. 
    In later iterations of the \BL\ algorithm more rectangles may be placed that intersect the line. 
    However, this will only increase the fraction of the line that is occupied by rectangles. 

    It remains to show that the statement holds for any horizontal line in the region~$H_i$.
    The~$k$ rectangles that intersect~$\ell$ have their bottom face below~$H_i$ by definition of the horizontal strip partition.
    And the~$k$ rectangles have their top faces above~$H_i$, as else~$r_i$ can be placed lower on top of such a rectangle, since the~$k$ rectangles each have width at least~$w_{r_i}$.
    Hence, any proper horizontal line in~$H_i$ intersects the~$k$ rectangles that intersect~$\ell$, and thus each such line satisfies the statements as desired.
\end{proof}

Observe that when rectangles are ordered by decreasing width, then every region satisfies the condition in Lemma~\ref{lemma:1/3_occupied_line}.
Additionally, in this case, Baker et al.~\cite{BCR1980} show that the leftmost part of a proper horizontal  line is always occupied by a rectangle.

\section{The 13/6-Approximation}
This section is devoted to proving that the \BottomLeftAlgorithm\ is a~$13/6$-approxi\-mation when placing the rectangles in the so-called~$\FQW$-ordering.
We start in Section~\ref{sec:FQW} by constructing this novel ordering that is based on the $\FQW$-partition of the rectangles, and establish several key properties of the partition.
After that, in Section~\ref{sec:setup_analysis}, we use the horizontal strip partition to derive lower bounds on the area occupied by rectangles in the \BL\ packing.
Most regions in this partition are at least half occupied by rectangles, though a few require special attention, being in total a little less than half occupied.
Section~\ref{sec:below_r_L} develops tools to distinguish different types of horizontal lines, which leads to the construction of a quadratic program to bound the maximum unoccupied area in the special regions.
This analysis culminates in the proof of Theorem~\ref{thm:main_13over6_theorem} in Section~\ref{sec:final_proof}, followed by final remarks on our approach in Section~\ref{sec:final_remark}.

\subsection{The Algorithm}\label{sec:FQW}
Baker, Coffman, and Rivest~\cite{BCR1980} showed that the \BottomLeftAlgorithm\ is a~$3$-approximation for \StripPacking\  when the rectangles are placed in order of decreasing width.
Moreover, they show that the analysis is tight by constructing a lower bound instance.
Their example consists of a so-called \textit{checkerboard} followed by a very tall rectangle with small width that is placed on top of the checkerboard construction.
The checkerboard is essentially half occupied by rectangles and half unoccupied, which alone would result in a lower bound of~$2$ on the approximation ratio.
However, the final tall rectangle enhances their lower bound to~$3$.
Therefore, we aim to place as many tall rectangles as possible first on the strip bottom.
Indeed this improves the approximation guarantee of the \BottomLeftAlgorithm.

With this in mind, we partition the rectangles~$\R$ into three sets, and use this partition to design an ordering in which the \BL\ algorithm packs the rectangles.
First of all, let~$\F\subseteq \R$ be a maximal set of the tallest possible rectangles that fit next to each other on the bottom of the strip.
To be more precise, go over the rectangles in order of decreasing height and add a rectangle to~$\F$ if the sum of widths inside~$\F$ remains less or equal to the strip width~$W$.
In case two rectangles have the same height, we do not care in which order they are considered. 
Second of all, let~$\W$ be the rectangles in~$\R\setminus \F$ with width greater than half of the strip width.
Last of all, define~$\Q = \R \setminus (\W\cup \F)$ to be the set of remaining rectangles.
We call the constructed partition of rectangles the~\textit{$\FQW$-partition}.
Pseudocode for computing the partition is given in Algorithm~\ref{alg:ordering}.

\begin{algorithm}
    \caption{ $\FQW$-partition of rectangles}
    \label{alg:ordering}
    \hspace*{\algorithmicindent} \textbf{Given}: \StripPacking\ instance~$(\R,W)$. \\
    \hspace*{\algorithmicindent} \textbf{Result}: Partition of $\R= \F \cup\Q \cup \W$.
    \begin{algorithmic}[1] 
        \State Let $\F = \emptyset$.
        \For{rectangle $r\in\R$ ordered by decreasing height}
        \If{$w_r + \sum_{f\in \F} w_f  \leq W$}
            \State Add~$r$ to~$\F$.
        \EndIf
        \EndFor
        \State Let $\W = \{r\in \R\setminus \F\mid w_r > \frac12 W\}$.
        \State Let $\Q = \R \setminus (\W\cup \F)$.
        \State \Return $\F,\Q,\W$.
    \end{algorithmic}
\end{algorithm}

An important property of the~$\FQW$-partition is that rectangles in~$\Q\cup\W$ are not too tall.
For a number~$h\in \mathbb{R}$, we denote by~$\F_{\geq h}$ all rectangles in the set~$\F$ of height at least~$h$, i.e.,~$\F_{\geq h} = \{f\in \F \mid h_f \geq h\}$.
\begin{lemma}\label{lem:upper_bound_h_r}
    Let~$\F,\Q,\W$ be the sets of the~$\FQW$-partition of a \StripPacking\ instance~$(\R,W)$ according to Algorithm~\ref{alg:ordering}.
    For every~$r\in \Q\cup\W$, we have~$h_r\leq \frac12 h_\text{OPT}$.
\end{lemma}
\begin{proof}
    For~$r \in \Q\cup\W$ we have $w_r+\sum_{f\in \F_{\geq h_r}} w_f >W$ by the definition of~$\F$.
    Hence there are at least two rectangles in~$\F_{\geq h_r}\cup \{r\}$ that cannot be next to each other in the optimal packing.
    As each such rectangle has height at least~$h_r$, it follows that~$2h_r$ is a lower bound on the height~$h_\text{OPT}$ of an optimum packing.
\end{proof}

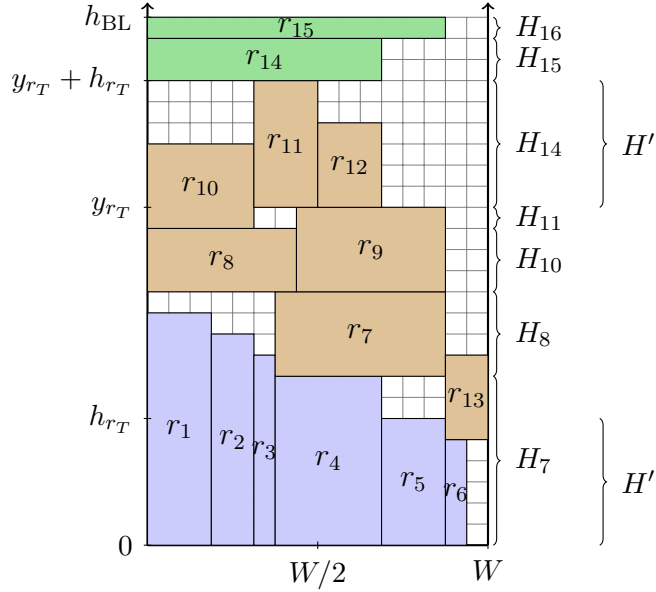
\begin{figure}[ht]
    \centering
    \begin{tikzpicture}[scale=0.38]
        % \strip{16}{27}
        \draw[help lines, very thin] (0,0) grid +(16,25);

        \draw[thick] (0,0) -- (16,0) node[anchor=north west] {};
        \draw[thick,->] (0,0) -- (0,25.7) node[anchor=south east] {};
        \draw[thick,->] (16,0) -- (16,25.7) node[anchor=south east] {};

        % mathcal{F}
       \filldraw[myfill]  (0,0) rectangle +(3,11); \draw ( 1.5, 5.5  ) node[] {$r_1$};
       \filldraw[myfill]  (3,0) rectangle +(2,10); \draw ( 4, 5  ) node[] {$r_2$};
       \filldraw[myfill]  (5,0) rectangle +(1, 9); \draw ( 5.5, 4.5  ) node[] {\small{$r_3$}};
       \filldraw[myfill]  (6,0) rectangle +(5, 8); \draw ( 8.5, 4  ) node[] {$r_4$};
       \filldraw[myfill] (11,0) rectangle +(3, 6); \draw ( 12.5, 3  ) node[] {$r_5$};
       \filldraw[myfill] (14,0) rectangle +(1, 5); \draw ( 14.5, 2.5 ) node[] {\small{$r_6$}};
        
        % mathcal{Q}
       \filldraw[mybrownfill] ( 6, 8) rectangle +(8,4);  \draw ( 10,10   ) node[] {$r_7$};
       \filldraw[mybrownfill] ( 0,12) rectangle +(7,3);  \draw ( 3.5,13.5) node[] {$r_8$};
       \filldraw[mybrownfill] ( 7,12) rectangle +(7,4);  \draw (10.5,14  ) node[] {$r_9$};
       \filldraw[mybrownfill] ( 0,15) rectangle +(5,4);  \draw ( 2.5,17  ) node[] {$r_{10}$};
       \filldraw[mybrownfill] ( 5,16) rectangle +(3,6);  \draw ( 6.5,19  ) node[] {$r_{11}$};
       \filldraw[mybrownfill] ( 8,16) rectangle +(3,4);  \draw ( 9.5,18  ) node[] {$r_{12}$};
       \filldraw[mybrownfill] (14, 5) rectangle +(2,4);  \draw (15  , 7  ) node[] {$r_{13}$};

        % mathcal{W}
       \filldraw[mygreenfill] (0,22) rectangle +( 11,2);  \draw (5.5  , 23  ) node[] {$r_{14}$};
       \filldraw[mygreenfill] (0,24) rectangle +(14,1);  \draw (7  , 24.5  ) node[] {{$r_{15}$}};

       \def\ytick#1#2{\draw (5pt,#1) -- (-5pt,#1) node[anchor=east] {\small #2};}
       \def\xtick#1#2{\draw (#1,5pt) -- (#1,-5pt) node[anchor=north] {\small #2};}

        \ytick{0}{$0$}
        % \ytick{5}{$h_{q_7}$}
        \ytick{6}{$h_{\rtop}$}
        % \ytick{8}{$y_{q_1}$}
        % \ytick{12}{$y_{q_2} =y_{q_3}$}
        % \ytick{15}{$y_{q_4}$}
        \ytick{16}{$y_{\rtop}$}
        \ytick{22}{$y_{\rtop}+h_{\rtop}$}
        \ytick{25}{$h_{\text{BL}}$}

        \xtick{8}{$W/2$}
        \xtick{16}{$W$}

        \def\rightbrace#1#2#3{\draw[decoration={brace,mirror,raise=2pt},decorate] (16,#1) -- node[right=6pt] {\small #3} (16,#2);}
        \def\farrightbrace#1#2#3{\draw[decoration={brace,mirror,raise=2pt},decorate] (21,#1) -- node[right=6pt] {\small #3} (21,#2);}

        \farrightbrace{0}{6}{$H'$}
        \rightbrace{0}{8}{$H_7$}
        \rightbrace{8}{12}{$H_8$}
        \rightbrace{12}{15}{$H_{10}$}
        \rightbrace{15}{16}{$H_{11}$}
        \rightbrace{16}{22}{$H_{14}$}
        \farrightbrace{16}{22}{$H'$}
        \rightbrace{22}{24}{$H_{15}$}
        \rightbrace{24}{25}{$H_{16}$}
    \end{tikzpicture}
    \caption{The packing~$\text{BL}(\mathcal{I}_\FQW)$ together with the horizontal strip partition.
    Blue rectangles are in~$\F$, brown in~$\Q$, and green in~$\W$.}
    \label{fig:FQW-BL_example}
\end{figure}

For a \StripPacking\ instance~$\mathcal{I}$, we order the rectangles based on the~$\FQW$-partition as follows: first sort the rectangles from~$\F$ by decreasing height, next order the rectangles from~$\Q$ by decreasing width, and finally take any ordering of~$\W$.
We break ties arbitrarily.
We denote the \BL\ packing obtained from this \textit{$\FQW$-ordering} by~$\text{BL}(\mathcal{I}_\FQW)$. 
Figure~\ref{fig:FQW-BL_example} illustrates an example.

\subsection{Setup of Analysis}\label{sec:setup_analysis}
We dedicate this part to the analysis of the approximation factor of the \BottomLeftAlgorithm\ following the~$\FQW$-ordering.
For the regions of the horizontal strip partition of this packing, we derive lower bounds on the fraction that is occupied by rectangles.
These bounds imply an approximation ratio of~$13/6$ as stated in Theorem~\ref{thm:main_13over6_theorem}.

We index the rectangles from the instance~$(\R,W)$ by~$r_1,\dots,r_n$ such that the \BL\ algorithm places~$r_i$ before~$r_j$ if~$i<j$.
Furthermore, we denote~$a = |\F|$ and~$b=|\Q|$, which implies that~$\F = \{r_1,\dots,r_a\}$,~$\Q=\{r_{a+1},\dots,r_{a+b}\}$ and~$\W=\{r_{a+b+1},\dots,r_n\}$ (see Figure~\ref{fig:FQW-BL_example}).

Let~$H_1,\dots,H_{n+1}$ be the horizontal strip partition as defined in Section~\ref{sec:horizontal_strip_partition}.
Observe that the regions~$H_1,\dots,H_a$ are empty, because all rectangles from~$\F$ are placed on the bottom of the strip.
Furthermore, in case that~$\Q=\varnothing$ or the highest top face of a rectangle from~$\Q$ is placed below height~$h_\text{max}$, then we even have a~$2$-approximation.
\begin{lemma}\label{lemma:simple_BL_FQW_2_apx_}
    If~$\Q=\varnothing$ or~$\max\{y_r+h_r\mid r\in \Q\} \leq h_\text{max}$, then the \BottomLeftAlgorithm\ of~$\mathcal{I}_\FQW$ has approximation ratio~$2$.
\end{lemma}
\begin{proof}
    All rectangles from~$\F$ are placed on the bottom of the strip and have height at most~$h_\text{max}$.
    If there are no rectangles in~$\Q$, or if the highest top face in~$\Q$ is below~$h_\text{max}$, then in the worst case all rectangles from~$\W$ are stacked on top of each other at height~$h_\text{max}$.
    Thus it holds that~$h_\text{BL} \leq h_\text{max} + \sum_{r\in\W} h_{r}$.
    Now both $h_{\max}$ and $\sum_{r\in\W} h_{r}$ are lower bounds on the height of an optimal packing, which implies~$h_\text{BL} \leq 2h_\text{OPT}$.
\end{proof}

Thus, from now on, we may assume that~$\Q\neq\varnothing$ and that there exists some rectangle~$r\in \Q$ for which it holds that~$y_r+h_r > h_\text{max}$.
The next result explains why the rectangles from~$\W$ are placed after rather than before~$\Q$.
\begin{lemma}\label{lemma:H_{a+1}}
    Every proper horizontal line in~$H_{a+1}$ is at least half occupied.
\end{lemma}
\begin{proof}
    The rectangles in~$\F$ are placed next to each other at the bottom of the strip and are ordered from left to right by decreasing height. 
    The first rectangle~$r_{a+1}$ from~$\Q$ is placed according to the \BL\ algorithm on top of this first row at height $y_{r_{a+1}}$. 
    Let~$f$ be the rightmost rectangle in~$\F$ whose top face is at height~$y_{r_{a+1}}$.
    Such a rectangle always exists as~$r_{a+1}$ must have a bottom supporter (cf. Section~\ref{sec:BL_algo}). 
    As~$r_{a+1}$ cannot be placed at a lower position, it follows that the gap between the right side of~$f$ and the right strip boundary is strictly less than~$w_{r_{a+1}}$,    which is less than~$\frac12 W$ by definition of~$\Q$. 
    Therefore, each horizontal line in~$H_{a+1}$ is at least half occupied. 
\end{proof}

We continue by analyzing the regions corresponding to rectangles from~$\W$.
\begin{lemma}\label{lemma:H_{n+1}}
    For~$a+b+2\leq i\leq n+1$, every proper horizontal line in the region~$H_i$ is at least half occupied.
\end{lemma}
\begin{proof}
    Each region~$H_i$ contains at least one rectangle from~$\W$.
    Thus, a proper horizontal line in~$H_i$ intersects a rectangle with width at least~$W/2$.
\end{proof}

Next, we consider the region~$H_{a+b+1}$ between the highest bottom face of a rectangle from~$\Q$ and the bottom face of the first rectangle that is placed from~$\W$.
Define the \textit{top rectangle of~$\Q$} as the rectangle~$\rtop$ in~$\Q$ whose top face is placed highest by the \BL\ algorithm; and in case of ties we let~$\rtop$ be one with highest bottom face 
(in the example shown in Figure~\ref{fig:FQW-BL_example} the rectangle $r_{11}$ is the rectangle $r_T$).
Then any proper horizontal line in~$H_{a+b+1}$ intersects~$\rtop$.
The rectangle~$\rtop$ might be the only rectangle that such a line intersects, and~$w_{\rtop}$ can be small, hence an arbitrarily small fraction of the line might be occupied by rectangles.
Thus, the region~$H_{a+b+1}$ can be very sparsely occupied. However, for a superset 
of~$H_{a+b+1}$ the following lemma shows that we get again an at least half occupied space. 
Note that we have~$H_{a+b+1} \subseteq [0,W]\times[y_{\rtop},y_{\rtop}+h_{\rtop}]$.
\begin{lemma}\label{lemma:space_H_prime}
   The space $H'=[0,W]\times ([0,h_{\rtop}]\cup [y_{\rtop},y_{\rtop}+h_{\rtop}])$ is at least half occupied. 
\end{lemma}
\begin{proof}
    It holds that~$y_{\rtop} \ge h_{\rtop}$ by construction of~$\F$.
    Moreover,  the total width of the rectangles in $\{\rtop\}\cup \F_{\geq h_{\rtop}}$ exceeds the width of the strip. 
    Thus the \BL\ algorithm occupies inside $H'$ an area of at least~$W\cdot h_{\rtop}$ by rectangles.
    As $H'$ has area $2 W\cdot h_{\rtop}$ it is at least half occupied.
\end{proof}

It remains to study the regions~$H_i$ for~$i\in \{a+2,\dots,a+b\}$.
For this, we define the \textit{left rectangle of~$\Q$} to be the first rectangle~$\rleft$ from $\Q$ that touches the left strip boundary
(in the example shown in Figure~\ref{fig:FQW-BL_example} the rectangle $r_{8}$ is the rectangle $r_L$).
If no rectangle in~$\Q$ touches the left strip boundary, then we set~$\rleft := \rtop$.
In Section~\ref{sec:below_r_L} we will develop a technique to prove that the union~$H_{a+2}\cup\cdots\cup H_{L}$ is almost half occupied by rectangles.
But first, we consider the regions~$H_{L+1},\dots,H_{a+b}$.
Since the rectangles in~$\Q$ are ordered by decreasing width we can apply Lemma~\ref{lemma:1/3_occupied_line}.
\begin{lemma}\label{lemma:H_{L+1}}
    For~$L+1 \leq i \leq a+b$, either~$H_i \subseteq H'$ or every proper horizontal line in the region~$H_i$ is at least half occupied.
\end{lemma}
\begin{proof}
    If $\rleft=\rtop$, then~$H_i\subseteq H'$.
    We thus may assume that $\rleft \not = \rtop$.
    A proper horizontal line~$\ell$ in~$H_i$ is above the bottom face of~$\rleft$.
    Therefore, prior to placing~$r_i$, the line~$\ell$ only intersects rectangles from~$\Q$ and these rectangles have width at least~$w_{r_i}$.
    Additionally, we will show that the leftmost part of~$\ell$ is occupied by a rectangle.
    Then Lemma~\ref{lemma:1/3_occupied_line} implies that~$\ell$ is at least half occupied.

    This argument is based on~\cite{BCR1980}.
    Let~$r'$ be the highest rectangle placed before~$r_i$ that touches the left strip boundary.
    Such~$r'$ exists and is part of~$\Q$, because~$\rleft \in\Q$ touches the left strip boundary and no rectangle from~$\F$ is above~$\rleft$.
    Since~$r'$ is placed before~$r_i$, it follows that the bottom face of~$r'$ must be placed below the proper line~$\ell$ by definition of the horizontal strip partition.
    If~$r'$ intersects~$\ell$, then the first part of~$\ell$ is occupied by a rectangle. 
    Otherwise, we show that there is no rectangle placed before~$r_i$ that is above~$r'$ and whose left face is strictly to the left of the right face of~$r'$.
    Namely, suppose there exists such a rectangle, then let~$r''$ be the leftmost such rectangle.    
    With the same argument as before, as~$\ell$ is a proper horizontal line in~$H_i$, the bottom face of~$r''$ is also below~$\ell$.
    Furthermore,~$r''$ must be supported on the left, but as~$r''$ is the leftmost rectangle, such a supporter is the left strip boundary, a contradiction to the definition of~$r'$.
    This implies that just before~$r_{i}$ is placed, the space above~$r'$ is unoccupied, and $w_{r'} \geq w_{r_{i}}$ by the decreasing width ordering of~$\Q$, hence~$r_{i}$ could have been placed lower on top of~$r'$ by the \BL\ algorithm.
    This contradiction implies that the leftmost part of the line~$\ell$ must be occupied by a rectangle.
\end{proof}

\subsection{The Space\texorpdfstring{~\boldmath $H_{a+2}\cup\cdots\cup H_L$}{ H-{a+2} cup H-{L}}}\label{sec:below_r_L}
A proper horizontal line in the regions~$H_{a+2},\dots,H_L$ might be less than half occupied prior to placing the first rectangle above the line, because, contrary to before, the leftmost part of the line is either occupied by rectangles from~$\F$ or is unoccupied.
However, if the line intersects~$k$ rectangles from~$\Q$ prior to placing the first rectangle above the line, then, according to Lemma~\ref{lemma:1/3_occupied_line}, the line is at least a~$k/(2k+1)\geq 1/3$-fraction occupied.
The central idea of this section is to establish a bound on the number of lines intersecting a fixed number of rectangles, which will subsequently imply that the union of the regions~$H_{a+2}\cup\cdots\cup H_{L}$  is occupied by at least the amount~$\frac12 hW - \frac{1}{12}h_\text{OPT} W$ where~$h=y_{\rleft} - y_{r_{a+1}}$ is the height of the union of these regions.

We begin with two lemmas that bound the height of any rectangle in~$\Q$, and consequently, the height of the space~$H_{a+2}\cup\cdots\cup H_{L}$.
For this, define~$\rbot$ to be the leftmost bottom supporter of~$r_{a+1}$ (cf. Section~\ref{sec:BL_algo}), that is, of the first rectangle that is placed from~$\Q$.
\begin{lemma}\label{lemma:height_bounded_by_bottom_supporter}
    For each  rectangle~$r \in \Q$, we have~$h_{r} \leq h_{\rbot}$.
\end{lemma}
\begin{proof}
    Since~$r_{a+1}$ is the first rectangle that is placed from~$\Q$, it follows that the bottom supporter~$\rbot$ is a rectangle from~$\F$.
    It follows that~$h_{r_{a+1}} \leq h_{\rbot}$, as otherwise~$r_{a+1}$ would have been part of~$\F$.
    As~$\Q$ is ordered by decreasing width, it holds for any rectangle~$r \in\Q$ that~$w_{r} \leq w_{r_{a+1}}$.
    Therefore, it must hold that~$h_{r}\leq h_{\rbot}$, as else~$r$ would belong to~$\F$.
\end{proof}

This yields an upper bound on the height at which the left rectangle~$\rleft$ can be placed.
\begin{lemma}\label{lemma:bound_h_r_left}
    It holds that~$y_{\rleft} \leq h_\text{max} + h_{\rbot}$.
\end{lemma}
\begin{proof}
    The statement is certainly true if $\rleft = r_{a+1}$. 
    Thus, we may assume that there is at least one rectangle in $\Q$ that is placed before $\rleft$; let~$r\in \Q$ be the leftmost such rectangle whose left face touches the right face of a rectangle from~$\F$.
    If there are multiple such rectangles, then let $r$ be the top one. 
    Observe that by Lemma~\ref{lemma:height_bounded_by_bottom_supporter} the top face of $r$ is at most at height~$h_{\max} + h_{\rbot}$. 
    There are two options. 
    If there exists a rectangle in $\Q$ that is placed to the left of $r$, then the first such rectangle must be $\rleft$ because the only available left supporter left of $r$ is the left strip boundary by definition of $r$. 
    Otherwise, the space left of $r$ is not occupied by rectangles from $\Q$. 
    Now, the first rectangle that is placed above $r$ (if it exists) must be placed on top of~$r$ and must touch the left strip boundary; because the width of such a rectangle is less or equal to the width of $r$. 
    In conclusion, either there is a first rectangle in $\Q$ touching the left strip boundary, which then is placed at height at most $h_{\max} + h_{\rbot}$; or there is no such rectangle, and then the bottom face of $\rleft=\rtop$ is placed below the top face of $r$, implying that $y_{\rleft} < h_{\max} + h_{\rbot}$. 
\end{proof}

Next, we study the properties of horizontal lines that intersect a fixed number of rectangles just before placing a rectangle above the line.
To this end, define the \textit{type} of a proper horizontal line~$\ell$ as the set~$T(\ell)$ of rectangles from~$\Q$ that intersect~$\ell$ prior to placing the first rectangle above~$\ell$.
If the cardinality of~$T(\ell)$ equals~$k$, then we say that the line~$\ell$ has \textit{order~$k$}.
Moreover, denote by~$\LM(\ell)$ the leftmost rectangle from~$T(\ell)$.

\begin{lemma}\label{lemma:left_supporter}
    Let~$\ell$ be a proper horizontal line in~$H_{a+2}\cup \cdots\cup H_L$.
    Then~$\LM(\ell)$ has a rectangle from~$\F$ as left supporter.
\end{lemma}
\begin{proof}
    Suppose not, then there is a rectangle from~$\Q$ that is the left supporter of~$\LM(\ell)$.
    Let~$r$ be such a left supporter with highest~$y_r$ and let~$r'$ be the first rectangle that is placed above~$\ell$.
    Then it holds that~$w_r\geq w_{r'}$.
    Prior to placing~$r'$, no rectangle intersects~$\ell$ to the left of~$\LM(\ell)$, and no rectangle is above~$\ell$.
    As~$\ell$ is a proper line and the space to the left of~$\LM(\ell)$ is larger than~$w_{r'}$, it follows that~$r'$ can be placed with its bottom face below~$\ell$ to the left of~$\LM(\ell)$, contradicting that~$r'$ is above~$\ell$.
\end{proof}

For two horizontal lines~$\ell$ and~$\ell'$, we denote~$\ell<\ell'$ when~$\ell$ is below~$\ell'$.
\begin{lemma}\label{lemma:decrease_widths}
    Let~$\ell<\ell'$ be proper horizontal lines.
    Let~$r\in\Q$ be a rectangle such that~$r\in T(\ell)$ and let~$r'\in\Q$ be a rectangle such that both~$r'\in T(\ell')$ and~$r'\not\in T(\ell)$.
    % $r' \cap \ell =\varnothing$.
    Then~$w_r\geq w_{r'}$. 
\end{lemma}
\begin{proof}
    Suppose that~$w_r<w_{r'}$, then the \BL\ algorithm places~$r'$ before~$r$.
    As~$r\in T(\ell)$ is placed before the first rectangle is placed above~$\ell$, it follows that~$r'$ must intersect~$\ell$.
    However, this contradicts~$r'\not\in T(\ell)$.
\end{proof}

Let~$\mathcal{L}^k$ be the set of all proper horizontal lines of order~$k$ inside of the space~$H_{a+2}\cup\cdots\cup H_L$.
Define~$\R^k = \{r \in \Q \mid \exists \ell\in \mathcal{L}^k : r=\LM(\ell)\}$, that is, the set of all rectangles in~$\Q$ that are the leftmost rectangle of some line of order~$k$.
Furthermore, we denote by~$\LM^k$ the leftmost rectangle from~$\R^k$, and if there are multiple such rectangles, then let~$\LM^k$ be the highest among them.
\begin{lemma}\label{lemma:half_occupied_lower_lines}
    Let~$\ell\in \mathcal{L}^k$. Then either~$\LM(\ell) = \LM^k$, or~$\ell$ is at least half occupied.
\end{lemma}
\begin{proof}
    Suppose that~$\LM(\ell)\neq \LM^k$.
    Lemma~\ref{lemma:left_supporter} states that each rectangle in~$\R^k$ has a rectangle from~$\F$ as left supporter.
    As the rectangles in~$\F$ are placed from left to right by decreasing height, it follows that~$\LM^k$ also has the highest bottom face among all the rectangles in~$\R^k$.
    Let~$\ell' \in \mathcal{L}^k$ be a line intersecting~$\LM^k$.

    Let~$T(\ell) = \{r_{i_1},\dots,r_{i_k}\}$ and~$T(\ell') = \{r_{i_1'},\dots,r_{i_k'}\}$ ordered from left to right.
    There can be at most~$k+1$ gaps on line~$\ell$ respectively~$\ell'$ just before the first rectangle is placed above the line, namely, at most one to the left of~$r_{i_1}$ (resp.~$r_{i_1'}$), at most one between each consecutive pair of rectangles from the type, and at most one to the right of~$r_{i_k}$ (resp.~$r_{i_k'}$).
    Denote these gaps from left to right (with possibly length~$0$) by~$g_0,\dots,g_k$ respectively~$g_0',\dots,g_k'$.
    We have that~$x_{r_{i_1'}} \leq x_{r_{i_1}} - g_0$, because~$x_{r_{i_1'}} \leq x_{r_{i_1}}$ and both have a rectangle from~$\F$ as left supporter by Lemma~\ref{lemma:left_supporter}.
    This implies that
    \begin{align*}
        \sum_{j=0}^k g_j + \sum_{j=1}^k w_{r_{i_j}} \leq \sum_{j=1}^k (g_j' + w_{r_{i_j'}}).
    \end{align*}
    Now, Lemma~\ref{lemma:decrease_widths} implies that $\sum_{j=1}^k w_{r_{i_j}}\geq\sum_{j=1}^k w_{r_{i_j'}}$.
    Thus it follows that
    \begin{align*}
        \sum_{j=0}^k g_j  \leq \sum_{j=1}^k g_j'.
    \end{align*}
    Moreover, we know that each gap on~$\ell'$ has length less than the width of any rectangle adjacent to the gap just after the first rectangle is placed above~$\ell'$.
    As those rectangles have width less than or equal to the rectangles intersecting~$\ell$, it follows that 
    \begin{align*}
        \sum_{j=0}^k g_j  \leq \sum_{j=1}^k g_j' \leq \sum_{j=1}^k w_{r_{i_j'}} \leq \sum_{j=1}^k w_{r_{i_j}}.
    \end{align*}
    Thus the line~$\ell$ is at least half occupied, which proves the first statement.
    Figure~\ref{fig:second_order_prime} illustrates this for~$k=2$. 
    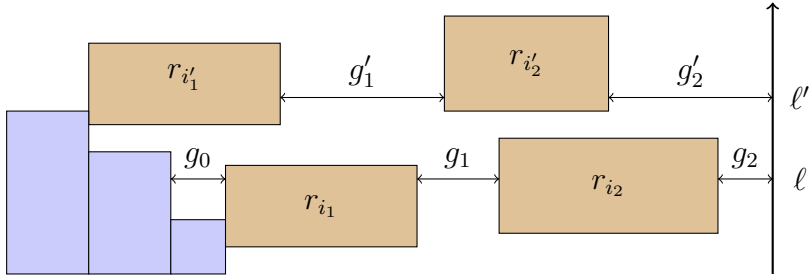
\begin{figure}
    \centering
    \begin{tikzpicture}[scale=0.18]
        % \strip{16}{27}
        \draw[thick,->] (50,0) -- (50,20) node[anchor=south east] {};

       \filldraw[mybrownfill] ( 0,11) rectangle (14,17);  \draw (7,14.5) node[] {$r_{i_1'}$};
       \filldraw[mybrownfill] ( 26,12) rectangle (38,19);  \draw (32,15.5) node[] {$r_{i_2'}$};

       \draw[<->] (14,13) -- (26,13) node[midway, above] {$g_1'$};
       \draw[<->] (38,13) -- (50,13) node[midway, above] {$g_2'$};

       \filldraw[myfill]  (-6,0) rectangle (0,12);
       % \draw (-3,6) node[] {$\F$};
       \filldraw[myfill]  (0,0) rectangle (6,9); 
       % \draw (3,4.5) node[] {$\F$};
       \filldraw[myfill]  (6,0) rectangle +(4,4); 
       % \draw (8,2) node[] {$\small \F$};
       \filldraw[mybrownfill] (10,2) rectangle (24,8);  \draw (17,5) node[] {$r_{i_1}$};
       \filldraw[mybrownfill] ( 30,3) rectangle (46,10);  \draw (38,6.5) node[] {$r_{i_2}$};

       \draw[<->] (6,7) -- (10,7) node[midway, above] {$g_0$};
       \draw[<->] (24,7) -- (30,7) node[midway, above] {$g_1$};
       \draw[<->] (46,7) -- (50,7) node[midway, above] {$g_2$};

        \draw (52,7) node[] {$\ell$};
        \draw (52,13) node[] {$\ell'$};

    \end{tikzpicture}
    \caption{Example showing that the lower horizontal line~$\ell$ of order~$2$ is at least half occupied. Blue rectangles are in~$\F$ and brown ones are in~$\Q$.}
    \label{fig:second_order_prime}
    \end{figure}
\end{proof}

\begin{corollary}
  \label{cor:i-line_occupation}
    Every~$\ell \in\mathcal{L}^k$ that intersects~$\LM^k$ is at least~${k}/{(2k+1)}$ occupied.
    All other lines of order~$k$ are at least~$1/2$ occupied.
\end{corollary}
\begin{proof}
    Consequences of Lemma~\ref{lemma:1/3_occupied_line} respectively Lemma~\ref{lemma:half_occupied_lower_lines}.
\end{proof}

\begin{lemma}\label{lemma:disjoint_downward_projections}
    Let~$\ell\in \mathcal{L}^k$,~$\ell'\in \mathcal{L}^{k'}$ and suppose that~$\ell<\ell'$.
    Then~$k\leq k'$.
    Moreover, if~$k<k'$, then the downward projection of the gap between the rightmost rectangle of~$\F$ that intersects~$\ell$ and~$\LM(\ell)$ is disjoint from the downward projection of the gap between the rightmost rectangle of~$\F$ that intersects~$\ell'$ and~$\LM(\ell')$. 
\end{lemma}
\begin{proof}
    Suppose that the first statement is false, then there exist~$\ell\in \mathcal{L}^k$ and~$\ell'\in\mathcal{L}^{k'}$ with~$\ell<\ell'$ and~$k>k'$, such that for every~$\ell''\in \mathcal{L}^{k''}$ with $\ell<\ell''<\ell'$ it holds that~$T(\ell'') = T(\ell)$ or~$T(\ell'')=T(\ell')$.
    This means that all rectangles in~$T(\ell')\setminus T(\ell)$ are placed on top of rectangles of~$\F\cup T(\ell)\setminus T(\ell')$.
    Consider the following assignment: go over the rectangles in~$r'\in T(\ell')\setminus T(\ell)$ from left to right, and assign a rectangle~$r\in T(\ell)\setminus T(\ell')$ to~$r'$ if~$r'$ is the first rectangle in this ordering of~$T(\ell')\setminus T(\ell)$ that is above~$r$.
    By definition, a rectangle from~$T(\ell')\setminus T(\ell)$ cannot be matched with two rectangles, because then it cannot have a left supporter as its width is less than the width of the rectangles from~$T(\ell)\setminus T(\ell')$ by Lemma~\ref{lemma:decrease_widths}.
    Thus each rectangle is matched with at most one other rectangle, but then as~$k > k'$, there must be a rectangle in~$T(\ell)\setminus T(\ell')$ that has no rectangle from~$T(\ell')\setminus T(\ell)$ on top of it, and therefore, the first rectangle that is placed above~$\ell'$ can now be placed lower on top of this free rectangle. Contradiction.

    Next suppose that~$k\neq k'$.
    Let~$r$ be the rightmost rectangle from~$\F$ intersecting~$\ell$.
    Then~$x_{\LM(\ell')} \leq x_{r}+w_r$, because~$\F$ is placed in order of decreasing height,~$\ell'$ is above~$\ell$ and both have a left supporter in~$\F$ by Lemma~\ref{lemma:left_supporter}.    
    This immediately implies that the downward projections are disjoint, since the first gap on~$\ell'$ ends at~$x_{\LM(\ell')}$, and the first gap on~$\ell$ begins at~$x_{r} + w_{r}$.
\end{proof}

To measure the distance between two horizontal lines, we denote by~$d(\ell,\ell')$ the difference in~$y$-coordinates.
\begin{theorem}\label{theorem:H_{a+2}}
        The union of the regions~$H_{a+2}\cup\cdots\cup H_L$ consists of at least~$\frac12 h W - \frac{1}{12} h_\text{OPT} W$ occupied space with~$h = y_{\rleft} - y_{r_{a+1}}$.
\end{theorem}
\begin{proof}
    By Corollary~\ref{cor:i-line_occupation} the space~$H_{a+2}\cup\cdots\cup H_L$ is maximally unoccupied when we have for all~$k$ that~$|\R^k| \leq 1$, because only lines of order~$k$ intersecting the leftmost rectangle of~$\R^k$ can be~$(k+1)/(2k+1)$ unoccupied, while other lines of order~$k$ are only at most half unoccupied.

    Define~$\alpha_k = \sup\{ d(\ell,\ell') \mid \ell,\ell'\in \mathcal{L}^k \text{ s.t. } \LM^k \in T(\ell)\cap T(\ell') \}$, that is, the largest distance between two lines of order~$k$ that both intersect the leftmost rectangle of~$\R^k$.
    If no line of order~$k$ exists, then define~$\alpha_k = 0$.
    From Lemma~\ref{lemma:half_occupied_lower_lines} it follows that all lines of order $k$ that are less than half occupied are considered in this supremum.
    It holds that~$\alpha_k \leq h_{\LM^k} \leq \frac12 h_\text{OPT}$ by Lemma~\ref{lem:upper_bound_h_r}.
    Furthermore, the height of the space~$H_{a+2}\cup\cdots\cup H_L$ is~$h$ by definition, thus~$\sum_{k=1}^\infty \alpha_k = h$.
    Notice that for an instance consisting of~$n$ rectangles, it holds that~$\alpha_k = 0$ for any~$k>n$, hence~$\sum_{k=1}^\infty \alpha_k$ equals the finite sum~$\sum_{k=1}^n \alpha_{k}$. 
    In conclusion, we have the linear constraints~$\alpha_k \leq \frac12 h_\text{OPT}$ for every~$k$ and~$\sum_{k=1}^\infty \alpha_k = h$.

    Next, consider~$\ell \in \mathcal{L}^k$ that is less than half occupied. 
    This line might intersect some rectangles~$r_1,\dots,r_z$ from~$\F$ and it intersects~$k$ rectangles~$r_{i_1},\dots,r_{i_k}$ from~$\Q$ prior to placing the first rectangle above~$\ell$ (see Figure~\ref{fig:beta_ell_definition}).
    We define~$x_\F^\ell = x_{r_z} + w_{r_z}$ if~$r_z$ exists, and otherwise we define~$x_\F^\ell =0$.
    By Lemma~\ref{lemma:gap_width} at this moment, each gap between rectangles has width strictly less than the width of the smallest rectangle from~$\Q$ that intersects the line $\ell$.
    Hence, there exists an~$x$-coordinate~$x_\ell\in [x_\F^\ell,x_{r_{i_1}}] $ such that the line $\ell$ is exactly half occupied to the right of~$x_\ell$.
    Define the length of the remaining gap on the line by~$\beta^\ell = x_\ell - x_\F^\ell$.
    \begin{figure}[ht]
    \centering
    \begin{tikzpicture}[scale=0.55]
        \draw[thick,->] (0,-0.5) -- (0,5.7) node[anchor=south east] {};
 %       \draw[thick] (0,0) -- (23,0) node[anchor=south east] {};
        \draw[thick,->] (23,-0.5) -- (23,5.7) node[anchor=south east] {};

       \filldraw[myfill]  (0,0) rectangle (2,5); \draw (1,2.4) node[] {$r_1$};
       \filldraw[myfill]  (2,0) rectangle (3.5,4); \draw (2.75,2.4) node[] {$\cdots$};
       \filldraw[myfill]  (3.5,0) rectangle (5.5,3); \draw (4.5,2.4) node[] {$r_z$};

      \filldraw[mybrownfill] (9,1) rectangle (13,3);  \draw (11,2.4) node[] {$r_{i_1}$};
      \filldraw[mybrownfill] (16,1) rectangle (20,3);  \draw (18,2.4) node[] {$r_{i_2}$};

       \draw[<->] (5.5,2.5) -- (7,2.5) node[midway, above] {$\beta^\ell$};
       \draw[] (0,1.8) -- (23,1.8);
       \draw[thick] (5.5,-0.1) -- (5.5,0.1);
       \draw[very thin, gray, dashed] (7,-0.1) -- (7,2.5);
       \draw[very thin, gray, dashed] (9,-0.1) -- (9,1.8);
       \draw[thick] (7,-0.1) -- (7,0.1);
       \draw[thick] (9,-0.1) -- (9,0.1);
        \draw (-1,1.9) node[] {$\ell$};
        \draw (5.5,-0.8) node[] {$x_\F^\ell$};
        \draw (7,-0.9) node[] {$x_\ell$};
        \draw (9,-1) node[] {$x_{r_{i_1}}$};
    \end{tikzpicture}
    \caption{Example of a horizontal line~$\ell$ of order~$2$. To the right of~$x_\ell$ the line~$\ell$ is exactly half occupied prior to placing the first rectangle above~$\ell$, the remaining gap between~$x_{\F}^\ell$ and~$x_\ell$ is defined as~$\beta^\ell$.}
    \label{fig:beta_ell_definition}
    \end{figure}
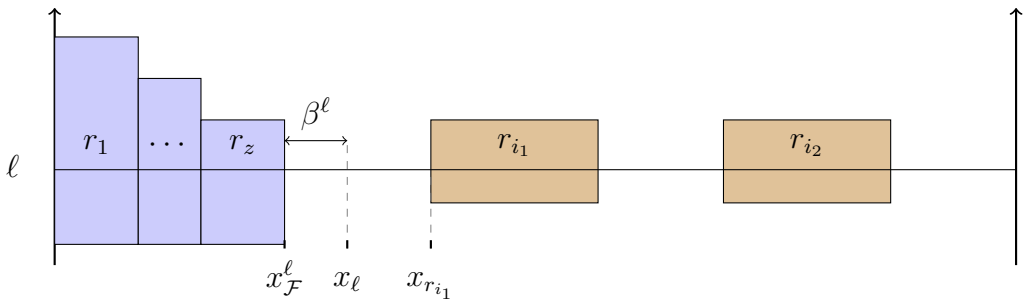
    
    Furthermore, for all lines of order~$k$ that are less than $W/2$ occupied,
    consider the largest such~$\beta^\ell$-gap
    \begin{align*}
        \beta_k = \max\{\beta^\ell\mid \ell\in\mathcal{L}^k\text{ is less than $W/2$ occupied}\}.
    \end{align*}
    Again define~$\beta_k = 0$ if~$\mathcal{L}^k$ is empty.
    It holds that the gap~$\beta_k$ is upper bounded by~$\frac{1}{2k+1}W$, because by Lemma~\ref{lemma:1/3_occupied_line} a line of order~$k$ is at most~$\frac{k}{2k+1}W$ occupied and each rectangle on the line is larger than any unoccupied gap on the line.
    Moreover, it holds that~$\sum_{k=1}^\infty \beta_k \leq W$, 
    because for different values of~$k$ the downward projections of the gaps defined by~$\beta_k$ are disjoint by Lemma~\ref{lemma:disjoint_downward_projections}.

    Next we describe the relation between the amount of unoccupied space and the values~$\alpha$ and~$\beta$.
    A line~$\ell$ of order~$k$ has exactly~$\beta^\ell + \frac12 \left(W - x_\ell\right)$ unoccupied space (see Figure~\ref{fig:beta_ell_definition}).
    By definition it holds that~$\beta^\ell \leq \beta_k$.
    Furthermore, we have~$x_\ell \geq \sum_{j=k}^\infty \beta_j$, because the rightmost point of the $\beta_{j+1}$-gap is left of the leftmost point of the~$\beta_j$-gap for all~$j$ by Lemma~\ref{lemma:disjoint_downward_projections}. 
    Therefore,~$\ell$ has at most~$\beta_k + \frac12 \left(W - \sum_{j=k}^\infty \beta_j\right)$ unoccupied space.
    At most~$\alpha_k$ of the lines of order~$k$ are more than half unoccupied by Lemma~\ref{lemma:half_occupied_lower_lines}.
    Hence the maximum amount of unoccupied space in~$H_{a+2}\cup\cdots\cup H_L$ is a solution to the following quadratic program
    \begin{align*}
        \max\quad\quad &\sum_{k=1}^\infty \left(\alpha_k \beta_k + \frac12 \alpha_k \left(W- \sum_{j=k}^\infty \beta_j\right) \right) \quad\text{subject to},\\
        & 0\leq \alpha_k \leq \frac12 h_\text{OPT} \quad\text{for all } k\in\mathbb{N}, \\
        & \sum_{k=1}^\infty \alpha_k = h, \\
        & 0\leq \beta_k \leq \frac{1}{2k+1}W \quad\text{for all } k \in\mathbb{N}, \\
        & \sum_{k=1}^\infty \beta_k \leq W.
    \end{align*}
    The objective of this quadratic program can be rewritten as
    \begin{align*}
        \max\quad\quad \frac12 &\sum_{k=1}^\infty \alpha_k \gamma_k \quad\text{where } \gamma_k = \beta_k + W - \sum_{j=k+1}^\infty \beta_j.
    \end{align*}
    Let~$k_1$ and~$k_2$ be different indices such that~$\gamma_{k_1}\geq \gamma_{k_2}\geq \gamma_k$ for all~$k\in \mathbb{N} \setminus \{k_1\}$.
    Then setting~$\alpha_{k_1} = \min\left\{ \frac12 h_\text{OPT}, h \right\}$, $\alpha_{k_2} = \max\left\{0, h - \frac12 h_\text{OPT} \right\}$ and~$\alpha_k = 0$ for~$k\in \mathbb{N}\setminus \{k_1,k_2\}$ maximizes the objective (provided that~$\gamma_{k_1},\gamma_{k_2} \geq 0$), because by Lemma~\ref{lemma:bound_h_r_left} we have that~$h =  y_{\rleft} - y_{r_{a+1}} \leq h_\text{max} \leq h_\text{OPT}$, as~$\rbot$ is the leftmost bottom supporter of~$r_{a+1}$.
    If~$h\leq \frac12 h_\text{OPT}$, then~$\alpha_{k_2} = 0$ and the maximum objective value is~$\frac12 h \gamma_{k_1} = \frac23 h W$ when~$k_1 = 1$, $\beta_1 = \frac13W$ and~$\beta_k = 0$ for all~$k\geq 2$.
    And observe that~$\frac23 h W = \frac12 hW + \frac16 hW \leq \frac12 hW + \frac{1}{12} h_\text{OPT}W$ as~$h \leq \frac12 h_\text{OPT}$.
    Otherwise, if~$h>\frac12 h_\text{OPT}$, then the objective is~$\frac14 h_\text{OPT} \gamma_{k_1} + \frac12 (h-\frac12 h_\text{OPT}) \gamma_{k_2}$.
    There exists a maximum solution with~$\beta_k = 0$ for~$k\not\in \{k_1,k_2\}$, because~$\beta_k>0$ only increases the objective when~$\alpha_k > 0$, which is only the case for~$k\in \{k_1,k_2\}$. 
    Thus if~$k_1<k_2$, then, as~$h\leq h_\text{OPT}$,  the objective becomes
    \begin{align*}
        \frac14 h_\text{OPT} (\beta_{k_1} + W - \beta_{k_2}) + \frac12 \left(h-\frac12 h_\text{OPT}\right) (\beta_{k_2} + W)
        &\leq \frac14 h_\text{OPT} \beta_{k_1} + \frac12 h W
    \end{align*}
    and in case $k_1 > k_2$, the objective is
    \begin{align*}
        \frac14 h_\text{OPT} (\beta_{k_1} + W ) + \frac12 \left(h-\frac12 h_\text{OPT}\right) (\beta_{k_2} + W - \beta_{k_1}) \leq  \frac14 h_\text{OPT} \beta_{k_2} + \frac12 hW.
    \end{align*}
     This is maximum for~$\min\{k_1,k_2\} = 1$ and~$\beta_{\min\{k_1,k_2\}} = \frac13W$.
    This also gives the desired result~$\frac{1}{12}h_\text{OPT}W + \frac12 hW$.
\end{proof}

\subsection{The Proof of the \texorpdfstring{\boldmath $13/6$}{13/6}-Approximation Ratio}
\label{sec:final_proof}
We now combine the above results to establish Theorem~\ref{thm:main_13over6_theorem}.

\thmmain*

\begin{proof}
    Consider the packing~$\text{BL}(\mathcal{I}_\FQW)$.
    If~$\Q=\varnothing$ or~$y_r+h_r \leq h_\text{max}$ for all~$r\in \Q$, then Lemma~\ref{lemma:simple_BL_FQW_2_apx_} implies that the \BL\ algorithm is a~$2$-approximation.
    Hence assume that~$\Q\neq\varnothing$ and~$\max\{y_r+h_r\mid r\in\Q\} > h_\text{max}$.
    The space~$H'$ defined in Lemma~\ref{lemma:space_H_prime} is contained in~$H_{a+1}\cup H_{L}\cup\cdots\cup H_{a+b+1}$ by Lemma~\ref{lemma:height_bounded_by_bottom_supporter}.
    Thus, Lemma~\ref{lemma:H_{a+1}},~$\ref{lemma:H_{n+1}}$,~\ref{lemma:space_H_prime} and~\ref{lemma:H_{L+1}} imply that the space~$H_{a+1}\cup H_{L}\cup\cdots\cup H_{a+b+1}$ is at least half occupied by rectangles.
    Furthermore, for~$h = y_{\rleft} - y_{r_{a+1}}$, Theorem~\ref{theorem:H_{a+2}} states that the space~$H_{a+2}\cup\cdots\cup H_L$ is at least~$\frac12 hW - \frac{1}{12}h_\text{OPT} W$ occupied.
    Thus the \BL\ algorithm covers at least an area of
    \begin{align*}
        \frac12 (h_\text{BL} - h)W + \frac12 hW - \frac{1}{12}h_\text{OPT} W=  \frac12 h_\text{BL} W - \frac{1}{12} h_{\text{OPT}} W.
    \end{align*}
    As the total area covered by rectangles divided by~$W$ is a lower bound for~$h_{\text{OPT}}$ we get that
    \begin{align*}
        h_{\text{OPT}} \ge  \frac12 h_\text{BL}  - \frac{1}{12} h_{\text{OPT}}.
    \end{align*}
     This implies the desired approximation guarantee
     \[  h_{\text{BL}}  \le \frac{13}{6} h_{\text{OPT}}.
%         \qedhere
     \]
\end{proof}

\subsection{Final Remarks on the\texorpdfstring{~\boldmath$\FQW$}{ FQW}-Ordering}
\label{sec:final_remark}
We conclude this section with remarks on the special case involving only squares, followed by lower bounds on the performance of our approach.
Moreover, we briefly discuss an alternative~$\FQW$-ordering of rectangles, where~$\F$ is ordered by increasing height, rather than decreasing height.

\subsubsection{The Square Case}
\label{sec:square_FQW}
In the special case where all rectangles are squares, the \BL\ algorithm on~$\mathcal{I}_\FQW$ has an approximation ratio of~$2$.
The reason for this is that packing~$\F \cup \Q$ using the~$\FQW$-ordering is identical to packing those squares in decreasing order of size, because placing the squares in~$\F\cup \Q$ by decreasing width coincides with placing them by decreasing height, thus the squares that are packed at the bottom of the strip are exactly the squares from~$\F$, and~$\Q$ is placed by decreasing size on top of~$\F$.
Baker et al.\cite{BCR1980} showed that a \BL\ packing of squares in decreasing size has at least half of its area occupied by squares. 
Adding further squares of width at least~$W/2$ on top preserves this property.
Consequently,~$h_\BL(\mathcal{I}_\FQW)\leq 2h_\text{OPT}(\mathcal{I})$ for squares.

\subsubsection{Lower Bounds}
The checkerboard instance of~\cite{BCR1980} gives a lower bound of~$2$ on the approximation ratio of the \BL\ algorithm when squares are packed in order of decreasing size.
The same lower bound holds for the~$\FQW$-ordering as the squares of size~$2-i\epsilon$ constitute the set~$\F$.

There remains a small gap between the lower bound of~$2$ and our upper bound of~$13/6$ for the approximation ratio of the \BL\ algorithm with~$\FQW$-ordering.
Closing this gap cannot be achieved merely by analyzing the horizontal strip partition and bounding the area occupied by rectangles in each region, as the following example demonstrates.

\begin{figure}[ht]
    \centering
    \begin{tikzpicture}[scale=0.45]
        % \strip{16}{27}
        \draw[help lines, very thin] (0,0) grid +(9,12.7);

        \draw[thick] (0,0) -- (9,0) node[anchor=north west] {};
        \draw[thick,->] (0,0) -- (0,12.7) node[anchor=south east] {};
        \draw[thick,->] (9,0) -- (9,12.7) node[anchor=south east] {};

        % mathcal{F}
       \filldraw[myfill]  (0,0) rectangle +(3,5); \draw ( 1.5, 2.5  ) node[] {$r_1$};
       \filldraw[myfill]  (3,0) rectangle +(4,4); \draw ( 5, 2  ) node[] {$r_2$};
        
        % mathcal{Q}
       \filldraw[mybrownfill] ( 3,4 ) rectangle +(4,4);  \draw ( 5, 6  ) node[] {$r_3$};
       \filldraw[mybrownfill] ( 0,8 ) rectangle +(4,1);  \draw ( 2, 8.5 ) node[] {$r_4$};
       \filldraw[mybrownfill] ( 4,8 ) rectangle +(4,4);  \draw ( 6, 10   ) node[] {$r_5$};

       \def\ytick#1#2{\draw (5pt,#1) -- (-5pt,#1) node[anchor=east] {\small #2};}
       \def\xtick#1#2{\draw (#1,5pt) -- (#1,-5pt) node[anchor=north] {\small #2};}

        \ytick{0}{$0$}
        \ytick{4}{$h$}
        \ytick{5}{$h+1$}
        \ytick{8}{$2h$}
        \ytick{9}{$2h+1$}
        \ytick{12}{$h_{\text{BL}}=3h$}

        \xtick{3}{}
        \draw (3,-0.8) node[] {\small $w$};
        \xtick{6}{}
        \draw (6,-0.8) node[] {\small $2w$};
        \xtick{9}{}
        \draw (10.5,-0.8) node[] {\small $W=3w$};

        % \xtick{9}{$W=3w$}
        
        \def\rightbrace#1#2#3{\draw[decoration={brace,mirror,raise=2pt},decorate] (9,#1) -- node[right=6pt] {#3} (9,#2);}

        \rightbrace{0}{4}{$H_3$}
        \rightbrace{4}{8}{$H_4$}
        \rightbrace{8}{12}{$H_6$}

    \end{tikzpicture}
    \caption{The \BL\ packing of an instance following the~$\FQW$-ordering on a strip of width~$W=3w$ with rectangles~$\F = \{(w,h+1),(w+1,h) \}$, ~$\Q = \{(w+1,h),(w+1,1) (w+1,h)\}$ and~$\W = \varnothing$.}
    \label{fig:FQW_lower_bound}
\end{figure}
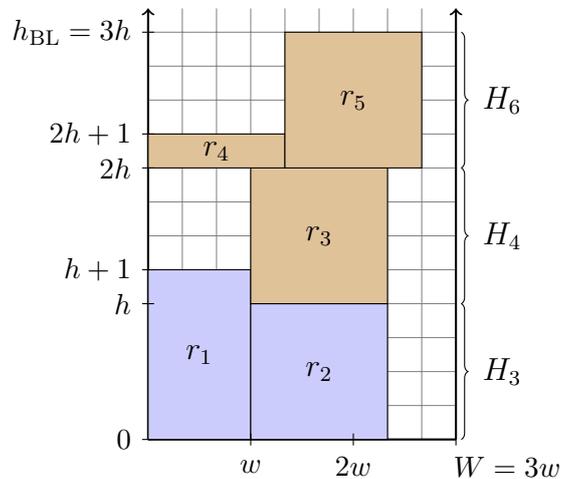

Consider the \BL\ packing in~$\FQW$-order for an instance with strip width~$3w$ consisting of one rectangle of size~$(w, h+1)$, three rectangles of size~$(w+1, h)$, and one of size~$(w+1, 1)$ (see Figure~\ref{fig:FQW_lower_bound}).
As~$h, w \to \infty$, the combined space~$H' = H_3 \cup H_6$ is half occupied, while~$H_4$ is only one-third occupied.
The total height of~$H_3 \cup H_6$ approaches~$h_\text{OPT}$, and the height of~$H_4$ approaches~$\frac12 h_\text{OPT}$.
Using the area bound this implies that
\begin{align*}
    \frac12 (h_\text{BL} - h_{r_3}) + \frac13 h_{r_3} = \frac12 h_\text{BL} - \frac16 h_{r_3} \leq h_\text{OPT}.
\end{align*}
This example shows that narrowing the gap in the approximation ratio of the \BottomLeftAlgorithm\ with~$\FQW$-ordering requires new techniques that analyze the union of the regions~$H_{a+2} \cup \cdots \cup H_L$ jointly with the other regions.
In particular, we have 
\begin{align*}
    h_\text{BL} \leq 2h_\text{OPT} + \frac13 h_{r_3} \leq \left(2 + \frac{1}{6}\right) h_\text{OPT}.
\end{align*}

\subsubsection{Ordering\texorpdfstring{~\boldmath$\F$}{ F} by Increasing Height}
An alternative natural ordering arises from the~$\FQW$-partition by arranging~$\F$ in order of increasing height (rather than decreasing), followed by sorting~$\Q$ and~$\W$ as before.
The advantage of this ordering is that every horizontal line in~$H_{a+2},\dots,H_{a+b}$ has its leftmost part occupied by a rectangle.
However, the drawback is that there may be a gap between the rightmost rectangle from~$\F$ and the right strip boundary, which complicates bounding the occupied area inside~$[0,W]\times[0,h_\text{max}]$.

Fortunately, using Lemma~\ref{lemma:1/3_occupied_line}, the regions~$H_{a+2},\dots,H_{a+b}$ that are above~$h_\text{max}$ are at least half occupied.
Also, Lemma~\ref{lemma:H_{n+1}} and~\ref{lemma:space_H_prime} still holds, hence the regions~$H_{a+b+2},\dots,H_{n+1}$ and the space~$H'$ are at least half occupied.
Now under the most pessimistic assumption, the space~$[0,W]\times[0,h_\text{max}]$ is completely unoccupied.
This results in an upper bound of~$3$ on the approximation ratio for the~\BL\ algorithm under this ordering of the rectangles, because with an area argument it holds that~$\frac12(h_\text{BL}-h_\text{max}) \leq h_\text{OPT}$. 

Whenever there is no gap between~$\F$ and the right strip boundary, we obtain a~$2$-approximation when ordering~$\F$ by increasing height, because, following the reasoning in Lemma~\ref{lemma:H_{a+1}}, the width of the unoccupied gap left of~$\F$ is at most the width of a rectangle from~$\Q$, which is bounded by~$W/2$.

\section{Conclusion}
In this paper, we presented a new ordering of rectangles under which the \BottomLeftAlgorithm\ achieves an approximation ratio of~$13/6$,  improving the previously best-known bound of~$3$ for the \BL\ algorithm given by~\cite{BCR1980}.
A key ingredient in our analysis is the detailed study of horizontal lines in the packing and the fraction of each line that is covered by rectangles.
For this, we developed a technique based on formulating and solving a quadratic program to determine the number of lines of a given order.
This method may also prove useful for refining the analysis of other packing algorithms.

Determining the exact approximation ratio of the \BL\ algorithm remains an open and intriguing problem.
There are instances for which no ordering of rectangles can achieve a ratio better than~$4/3-\epsilon$ for the \BL\ algorithm~\cite{hougardy2024bottom}, even when all rectangles are squares.
Closing the gap between~$4/3-\epsilon$ and~$13/6$ is interesting, because of the possibility for the \BL\ algorithm to surpass the current best-known~$(5/3+\epsilon)$-approximation for \StripPacking~\cite{harren2014}.

\section{Appendix}
In this appendix we show that the \BottomLeftAlgorithm\ has unbounded approximation ratio for a large class of rectangle orderings, including those based on area or diagonal length.
In particular, we construct a parameterized lower bound instance and prove that, for any ordering determined by a function~$f$ satisfying certain conditions, the \BL\ algorithm achieves a corresponding lower bound on its approximation factor.
After establishing this lower bound theorem, we illustrate its effectiveness in showing that many orderings have unbounded approximation ratio.
For example, this occurs when rectangles are placed according to their area, diagonal length, or any linear combination~$aw_r + bh_r$ of their width and height, where~$a,b > 0$.
\begin{theorem}\label{thm:general_BL_lower_bound}
    Let~$f\colon (0,W]\times\mathbb{R}_{>0}\to \mathbb{R}_{>0}$ be a function.
    Suppose that there exists a sequence~$h_1<\cdots<h_k$ in $\mathbb{R}_{>0}$ and~$y\in \mathbb{R}_{>0}$ such that the inverses~$(f_{h_i})^{-1}(y)$ exist.
    And assume that for~$W = \sum_{i=1}^k (f_{h_i})^{-1}(y)$, the inverse~$(f_W)^{-1}(y)$ also exists.
    Then the approximation ratio of the \BottomLeftAlgorithm\ when placing rectangles in any order of~$f$-value is at least
    \begin{align*}
        \frac{h_{\text{BL}}^f}{h_{\text{OPT}}} \geq \frac{\sum_{i=1}^{k} h_i + k\cdot (f_W)^{-1}(y)}{h_k + k\cdot (f_W)^{-1}(y)}.
    \end{align*}
\end{theorem}
\begin{proof}
    Consider a strip of width~$W = \sum_{i=1}^k (f_{h_i})^{-1}(y)$ together with the rectangles~$r_i = ((f_{h_i})^{-1}(y), h_i)$ and~$q_i = (W, (f_W)^{-1}(y))$  for~$1\leq i\leq k$.
    The~$f$-value of any of these rectangles equals~$y$, hence for any ordering of the rectangles based solely on the~$f$-values, the \BL\ algorithm cannot distinguish these rectangles.
    Therefore, consider the ordering~$r_1,q_1,\dots,r_k,q_{k}$ that alternatingly places~$r$ and~$q$ rectangles by increasing~$i$ index.
    As the height of the rectangles~$r_1,\dots,r_k$ is strictly increasing, and the rectangles~$q_i$ have width~$W$, it follows that~$r_i$ is placed on top of~$q_{i-1}$, and~$q_i$ is placed on top of~$r_i$.
    As all the rectangles are placed on top of each other, it follows that the height of the \BL\ packing is the sum of the heights of all the rectangles.

    An optimum packing places the rectangles~$q_1,\dots,q_k$ on top of each other, followed by placing the rectangles~$r_1,\dots,r_k$ next to each other, which is possible as the sum of their widths is exactly the width of the strip~$W$.
    Thus the height of an optimum packing equals the sum of the height of the rectangles~$q_1,\dots,q_k$ and the height~$h_k$ as this is the largest of the height amongst the rectangles~$r_1,\dots,r_k$.
\end{proof}

The first consequence of Theorem~\ref{thm:general_BL_lower_bound} is that ordering rectangles by area results in an unbounded approximation ratio of the \BL\ algorithm.

\begin{corollary}\label{cor:unbounded_area_order}
    The \BottomLeftAlgorithm\ has unbounded approximation ratio when solely basing the order of the rectangles on the values of the function~$f(w,h) = w^p h^q$ for~$p,q>0$.
    In particular, this holds for ordering the rectangles by area.
\end{corollary}
\begin{proof}
    Define~$\alpha = \frac{p}{q}$ and let~$h_i = i^\alpha$ and~$y=1$.
    Then~$(f_{h_i})^{-1}(1) = \frac{1}{i}$, the strip width is~$W = H_k$ (the~$k$-th harmonic number) and~$(f_W)^{-1}(1) = \frac{1}{H_k^\alpha}$.
    By Theorem~\ref{thm:general_BL_lower_bound} we have the lower bound 
    \begin{align*}
        \frac{h_{\text{BL}}^f}{h_{\text{OPT}}} &\geq \frac{\sum_{i=1}^{k} i^\alpha + \frac{k}{H_k^\alpha}}{k^\alpha + \frac{k}{H_k^\alpha}} \geq \frac{H_k^\alpha \int_0^k x^\alpha \ dx + k}{H_k^\alpha \cdot k^\alpha + k} = \frac{\frac{1}{\alpha+1} H_k^\alpha \cdot k^{\alpha+1} + k}{H_k^\alpha \cdot k^\alpha + k}.
    \end{align*}
    This is unbounded for~$k\to\infty$.
    For~$p=q=1$ the function~$f$ describes the area of a rectangle.
\end{proof}

The \BL\ algorithm is a~$3$-approximation when rectangles are ordered by decreasing width~\cite{BCR1980}.
The following corollary is surprising, because it states that when we order the rectangles by decreasing width plus a tiny fraction of the height (e.g., by decreasing~$w+10^{-100}\cdot h$ value), then the approximation ratio of the~\BL\ algorithm becomes unbounded.
\begin{corollary}
    The \BottomLeftAlgorithm\ has unbounded approximation ratio when solely basing the order of the rectangles on the values of the function~$f(w,h) = aw + bh$ with~$a,b>0$.
\end{corollary}
\begin{proof}
    For~$f(w,h) = aw+bh$, let~$h_i = a (1+i\epsilon)$,~$y = ab (1+k\epsilon)$ and choose~$\epsilon=\frac{2}{k(k-1)}$.
    Then we have~$(f_{h_i})^{-1}(y) = b(k-i)\epsilon$, the width of the strip is~$W = b$ and~$(f_W)^{-1}(y) = ak\epsilon$.
    By Theorem~\ref{thm:general_BL_lower_bound} a lower bound on the approximation ratio is
    \begin{align*}
        \frac{h_{\text{BL}}^f}{h_{\text{OPT}}} &\geq \frac{\sum_{i=1}^{k} a(1+i\epsilon) + k\cdot \frac{2a}{k-1}}{a(1+k\epsilon) + k\cdot \frac{2a}{k-1}} = \frac{k + \epsilon\frac{k(k+1)}{2} + \frac{2k}{k-1}}{1+\frac{2}{k-1}+ \frac{2k}{k-1}} \\
        &= \frac{k(k-1) + k+1 + 2k}{k-1 + 2 + 2k} = \frac{k^2+2k+1}{3k+1}.
    \end{align*}
    This is unbounded for~$k\to\infty$.
\end{proof}

\begin{corollary}
    The \BottomLeftAlgorithm\ has unbounded approximation ratio when solely basing the order of the rectangles on the length of their diagonal.
\end{corollary}
\begin{proof}
    For~$f(w,h) = \sqrt{w^2 + h^2}$, define~$h_i = \sqrt{1+i\epsilon}$,~$y = \sqrt{1+k\epsilon}$ and choose~$\epsilon = (\sum_{i=1}^{k-1}\sqrt{i})^{-2}$.
    Then we have~$(f_{h_i})^{-1}(y) = \sqrt{(k-i)\epsilon}$, the width of the strip is~$W=\sqrt{\epsilon}\sum_{i=1}^k \sqrt{k-i}=1$ and~$(f_W)^{-1}(y) = \sqrt{k\epsilon}$.
    By Theorem~\ref{thm:general_BL_lower_bound} it follows that 
    \begin{align*}
        \frac{h_{\text{BL}}^f}{h_{\text{OPT}}} &\geq \frac{\sum_{i=1}^{k} \sqrt{1+i\epsilon} + k\cdot \sqrt{k\epsilon}}{\sqrt{1+k\epsilon} + k\cdot \sqrt{k\epsilon}} \geq \frac{k\sqrt{1} + 0}{\sqrt{1+1/k}+ 1/\sqrt{k}} = \frac{k\sqrt{k}}{\sqrt{k+1}+1} \\
    \end{align*}
    because~$\epsilon \leq k^{-2}$.
    This is unbounded for~$k\to\infty$.
\end{proof}

\bibliographystyle{splncs04}

\end{document}